
%
%






\newcount\refnumber
\newcount\temp
\newcount\test
\newcount\tempone
\newcount\temptwo
\newcount\tempthr
\newcount\tempfor
\newcount\tempfiv
\newcount\testone
\newcount\testtwo
\newcount\testthr
\newcount\testfor
\newcount\testfiv
\newcount\itemnumber
\newcount\totalnumber
\refnumber=0
\itemnumber=0
\def\initreference#1{\totalnumber=#1
                 \advance \totalnumber by 1
                 \loop \advance \itemnumber by 1
                       \ifnum\itemnumber<\totalnumber
                        \temp=100 \advance\temp by \itemnumber
                        \count\temp=0 \repeat}

\def\ref#1{\temp=100 \advance\temp by #1
   \ifnum\count\temp=0
    \advance\refnumber by 1  \count\temp=\refnumber \fi
   \ [\the\count\temp]}

\def\reftwo#1#2{\tempone=100 \advance\tempone by #1
   \ifnum\count\tempone=0
   \advance\refnumber by 1  \count\tempone=\refnumber \fi
   \temptwo=100 \advance\temptwo by #2
   \ifnum\count\temptwo=0
   \advance\refnumber by 1  \count\temptwo=\refnumber \fi
 \testone=\count\tempone \testtwo=\count\temptwo
 \sorttwo\testone\testtwo
     \ [\the\testone,\the\testtwo]}       

\def\refthree#1#2#3{\tempone=100 \advance\tempone by #1
   \ifnum\count\tempone=0
    \advance\refnumber by 1  \count\tempone=\refnumber \fi
    \temptwo=100 \advance\temptwo by #2
   \ifnum\count\temptwo=0
    \advance\refnumber by 1  \count\temptwo=\refnumber \fi
    \tempthr=100 \advance\tempthr by #3
   \ifnum\count\tempthr=0
    \advance\refnumber by 1  \count\tempthr=\refnumber \fi
 \testone=\count\tempone \testtwo=\count\temptwo \testthr=\count\tempthr
 \sortthree\testone\testtwo\testthr
   \test=\testthr  \advance\test by -2
 \ifnum\test=\testone    \test=\testtwo  \advance\test by -1
    \ifnum\test=\testone   
    \ [\the\testone--\the\testthr]\fi \advance\temptwo by 1
  \else
     \ [\the\testone,\the\testtwo,\the\testthr]    
 \fi}

\def\reffour#1#2#3#4{\tempone=100 \advance\tempone by #1
   \ifnum\count\tempone=0
    \advance\refnumber by 1  \count\tempone=\refnumber \fi
    \temptwo=100 \advance\temptwo by #2
   \ifnum\count\temptwo=0
    \advance\refnumber by 1  \count\temptwo=\refnumber \fi
    \tempthr=100 \advance\tempthr by #3
   \ifnum\count\tempthr=0
    \advance\refnumber by 1  \count\tempthr=\refnumber \fi
    \tempfor=100 \advance\tempfor by #4
   \ifnum\count\tempfor=0
    \advance\refnumber by 1  \count\tempfor=\refnumber \fi
 \testone=\count\tempone \testtwo=\count\temptwo \testthr=\count\tempthr
 \testfor=\count\tempfor
 \sortfour\testone\testtwo\testthr\testfor
   \test=\testthr \advance\test by -1
   \ifnum\testtwo=\test   \test=\testtwo \advance\test by -1
    \ifnum\testone=\test  \test=\testfor \advance\test by -3
     \ifnum\testone=\test \ [\the\testone--\the\testfor]
     \else \ [\the\testone--\the\testthr,\the\testfor]
     \fi
    \else  \test=\testfor \advance\test by -1
     \ifnum\testthr=\test \ [\the\testone,\the\testtwo--\the\testfor]
     \else\ [\the\testone,\the\testtwo,\the\testthr,\the\testfor]
     \fi
    \fi
   \else \ [\the\testone,\the\testtwo,\the\testthr,\the\testfor]
   \fi}

\def\reffive#1#2#3#4#5{\tempone=100 \advance\tempone by #1
   \ifnum\count\tempone=0
    \advance\refnumber by 1  \count\tempone=\refnumber \fi
    \temptwo=100 \advance\temptwo by #2
   \ifnum\count\temptwo=0
    \advance\refnumber by 1  \count\temptwo=\refnumber \fi
    \tempthr=100 \advance\tempthr by #3
   \ifnum\count\tempthr=0
    \advance\refnumber by 1  \count\tempthr=\refnumber \fi
    \tempfor=100 \advance\tempfor by #4
   \ifnum\count\tempfor=0
    \advance\refnumber by 1  \count\tempfor=\refnumber \fi
    \tempfiv=100 \advance\tempfiv by #5
   \ifnum\count\tempfiv=0
    \advance\refnumber by 1  \count\tempfiv=\refnumber \fi
 \testone=\count\tempone \testtwo=\count\temptwo \testthr=\count\tempthr
 \testfor=\count\tempfor \testfiv=\count\tempfiv
 \sortfive\testone\testtwo\testthr\testfor\testfiv
  \test=\testthr \advance\test by -1
  \ifnum\testtwo=\test   \test=\testtwo \advance\test by -1
   \ifnum\testone=\test  \test=\testfor \advance\test by -3
    \ifnum\testone=\test \test=\testfiv \advance\test by -4
     \ifnum\testone=\test\ [\the\testone--\the\testfiv]
     \else\ [\the\testone--\the\testfor,\the\testfiv]
     \fi
    \else \ [\the\testone--\the\testthr,\the\testfor,\the\testfiv]
    \fi
   \else  \test=\testfor \advance\test by -1
    \ifnum\testthr=\test \test=\testfiv \advance\test by -2
     \ifnum\testthr=\test \ [\the\testone,\the\testtwo--\the\testfiv]
     \else \ [\the\testone,\the\testtwo--\the\testfor,\the\testfiv]
     \fi
    \else\ [\the\testone,\the\testtwo,\the\testthr,\the\testfor,\the\testfiv]
    \fi
   \fi
  \else \test=\testfor \advance\test by -1
   \ifnum\testthr=\test \test=\testfiv \advance\test by -2
    \ifnum\testthr=\test\
[\the\testone,\the\testtwo,\the\testthr--\the\testfiv]
    \else\ [\the\testone,\the\testtwo,\the\testthr,\the\testfor,\the\testfiv]
    \fi
   \else\ [\the\testone,\the\testtwo,\the\testthr,\the\testfor,\the\testfiv]
   \fi
  \fi}

\def\refitem#1#2{\temp=#1 \advance \temp by 100 \setbox\count\temp=\hbox{#2}}

\def\sortfive#1#2#3#4#5{\sortfour#1#2#3#4\relax
   \ifnum#5<#4\relax \test=#5\relax #5=#4\relax
     \ifnum\test<#3\relax #4=#3\relax
       \ifnum\test<#2\relax #3=#2\relax
         \ifnum\test<#1\relax  #2=#1\relax  #1=\test
         \else #2=\test \fi
       \else #3=\test \fi
     \else #4=\test \fi \fi}

\def\sortfour#1#2#3#4{\sortthree#1#2#3\relax
    \ifnum#4<#3\relax \test=#4\relax #4=#3\relax
       \ifnum\test<#2\relax #3=#2\relax
          \ifnum\test<#1\relax #2=#1\relax #1=\test
          \else #2=\test \fi
       \else #3=\test \fi \fi}

\def\sortthree#1#2#3{\sorttwo#1#2\relax
       \ifnum#3<#2\relax \test=#3\relax #3=#2\relax
          \ifnum\test<#1\relax #2=#1\relax #1=\test
          \else #2=\test \fi \fi}

\def\sorttwo#1#2{\ifnum#2<#1\relax \test=#2\relax #2=#1\relax #1=\test \fi}


\def\setref#1{\temp=100 \advance\temp by #1
   \ifnum\count\temp=0
    \advance\refnumber by 1  \count\temp=\refnumber \fi}

\def\printreference{\totalnumber=\refnumber
           \advance\totalnumber by 1
           \itemnumber=0
           \loop \advance\itemnumber by 1  
                 \ifnum\itemnumber<\totalnumber
                 \item{[\the\itemnumber]} \unhbox\itemnumber \repeat}
 \magnification=1200
  \hsize=15.5 truecm
  \vsize=23.0truecm
  \topskip=20pt            
  \fontdimen1\tenrm=0.0pt  
  \fontdimen2\tenrm=4.0pt  
  \fontdimen3\tenrm=7.0pt  
  \fontdimen4\tenrm=1.6pt  
  \fontdimen5\tenrm=4.3pt  
  \fontdimen6\tenrm=10.0pt 
  \fontdimen7\tenrm=2.0pt  
  \baselineskip=17.0pt plus 1.0pt minus 0.5pt  
  \lineskip=1pt plus 0pt minus 0pt             
  \lineskiplimit=1pt                           
  \parskip=2.5pt plus 5.0pt minus 0.5pt
  \parindent=15.0pt
\font\subsection=cmbx10 scaled\magstep1
\font\section=cmbx10 scaled\magstep 2
\font\bx=cmr8
\def\lsim{\; \raise0.3ex\hbox{$<$\kern-0.75em\raise-1.1ex\hbox{$\sim$}}\; }
\def\gsim{\; \raise0.3ex\hbox{$>$\kern-0.75em\raise-1.1ex\hbox{$\sim$}}\; }
\def\jump{\vskip 1truecm}
\def\GeV{\rm GeV}

\def\Re{{\rm Re}\hskip2pt}
\def\Im{{\rm Im}\hskip2pt}

\def\loota{{\scriptstyle\sqcap\kern-0.55em\hbox{$\scriptstyle\sqcup$}}}
\def\Loota{{\sqcap\kern-0.65em\hbox{$\sqcup$}}}

\def\1{\aa}
\def\vek#1{{\rm\bf #1}}

\def\inv#1{\frac{1}{#1}}
\def\del{\partial}
\def\slask{\!\!\!/}

\def\kk{{\rm k}}
\def\pp{{\rm p}}
%
%

\def\Abs#1{\left|#1\right|}

%

\def\np#1#2#3{{\it  Nucl.\ Phys.\ }{{\bf #1} {(#2)} {#3}}}
\def\pr#1#2#3{{\it Phys.\ Rev.\ }{{\bf #1} {(#2)} {#3}}}

\def\rmp#1#2#3{{\it  Rev.\ Mod.\ Phys.\ }{{\bf #1}{(#2)}{#3}}}

\def\zp#1#2#3{{\it Z.\ Phys.\ }{{\bf #1} {(#2)} {#3}}}

\def\frac#1#2{{#1\over#2}}
\newcount\eqnumber
\eqnumber=1
\def\chaphead{}

\def\new{\hbox{(\chaphead\the\eqnumber}\global\advance\eqnumber by 1}
\def\eqref#1{\advance\eqnumber by -#1 (\chaphead\the\eqnumber
     \advance\eqnumber by #1 }
\def\first{\hbox{(\chaphead\the\eqnumber{a}}\global\advance\eqnumber by 1}
\def\last{\advance\eqnumber by -1 \hbox{(\chaphead\the\eqnumber}\advance
     \eqnumber by 1}
\def\eq#1{\advance\eqnumber by -#1 equation (\chaphead\the\eqnumber
     \advance\eqnumber by #1}
\def\eqnam#1{\xdef#1{\chaphead\the\eqnumber}}


\def\eqt#1{Eq.~({{#1}})}

\def\eqand#1#2{Eqs.~({$#1$}) \rm and ({$#2$})}
\centerline {\section 1. Introduction}
\jump

First order electroweak phase transition  has often been discussed\ref{21}
within the
context of the finite temperature effective potential of the order parameter,
which is the expectation value of the Higgs field, evolving
in a thermal background. Such phase transitions
are highly local phenomena:
critical bubbles have a  finite size. However, effective potentials are
averaged over all space and do not describe
 local fluctuations, which may affect the
dynamics of nucleation and bubble growth. Fluctuations with spatial
correlations comparable to the critical
bubble size may be expected to be important for bubble nucleation
and, therefore, we concentrate on the thermalization of an
on-shell Higgs field with zero spatial momentum
($p_0=m_H,\ \vek{p}=0$). This choice reflects the fact that critical bubbles
are typically much bigger than the inter particle distance $\sim 1/T$ in
plasma.

Local fluctuations of the order parameter, which one might call subcritical
fluctuations, are continuously created and
destroyed by the self--interactions of the thermal background. A natural way
to describe them is to consider a non--constant order
parameter, for which
one needs the complete effective action for the background field $\phi(x)$.
The computation of the effective action is in general very difficult, but if
one focuses on small--amplitude fluctuations, $\phi(x)=\phi_0+b(x)$ with
$b(x)$ small, one may make use of a perturbation expansion in $b(x)$, and to
order $b^2$ the finite temperature effective action reads simply\ref{20}
\eqnam{\action}
$$
\Gamma[\phi_0,b]=
-V_{eff}(\phi_0)+\Gamma^{(1)}(\phi_0)b(k=0)
+\frac 12\int{d^4k\over (2\pi)^4}b(-k)\Gamma^{(2)}(\phi_0;k)b(k)\ ,
\eqno\new)
$$
where the tadpole $\Gamma^{(1)}=0$ at the high temperature minimum $\phi_0=0$,
and $\Gamma^{(2)}$ is the two--point function for the Higgs field, which
can be computed in ordinary perturbation theory.

Two--loop dispersion relations and the evolution of the
Fourier modes $b(k)$ in
$\lambda\phi^4$--theories has been discussed in detail in\ref{20}. In the
present
paper we investigate how fast the fluctuations thermalize in the Standard
Model above, and just prior to, the electroweak phase transition.
We assume the standard particle content and one Higgs doublet.
The thermalization
rate $\gamma$ depends on the imaginary part of the effective
action $\Im\Sigma \equiv -\Im\Gamma^{(2)}$, via
\eqnam{\thermalrate}
$$
\gamma=-{\Im \Sigma\over \omega},
\eqno\new)
$$
where $\omega=\omega (k)$ is the energy of a given mode. We compute
the imaginary part of the two--point function at one--loop level for fermion
and gauge boson corrections (at one loop there is no
higgs contribution to $\Im\Sigma$ in the unbroken phase). The
imaginary parts are cuts of the relevant one--loop
diagrams\reftwo{15}{3}, and therefore
equivalent to processes which may be called  decay, inverse decay,
absorption, or emission of
the higgs. For massless fermions and gauge bosons these processes are
kinematically forbidden at one--loop level. However, the physical
external states should include thermal corrections which, to leading order in
$T$,
are the hard thermal loops. They merely change the pole structure of the
propagators to account for
the interactions with the thermal background.
Accordingly, there appear energy thresholds, which in particular
for the fermionic loops are rather complicated because of the complicated
nature of the dispersion relations. Here, the one--loop case is exceptional
in that at higher loops there are no thresholds.

The paper is organized as follows. Section 2 contains a discussion about
the fermion dispersion relations at finite temperature and different
channels, and we present a derivation of the
fermionic contribution to the thermalization rate. In Section 3 we compute
the gauge boson loop and estimate also the size of the two--loop corrections.
In Section 4 we compare the thermalization rate to the electroweak phase
transition rate, and
Section 5 contains our conclusions together with a summary of our results.
\vfill\eject

\centerline {\section 2. Fermions}
\jump
\centerline{\subsection 2.1 Preliminaries}
\jump

The dispersion relation of massless chiral fermions is
drastically changed by the high temperature correction\reftwo{1}{2}. The
leading $T^2$ part of the fermion
self-energy gives rise to two kinds of excitations of the
thermal heat bath called particles and holes\ref{5}. Both particles and holes
have unusual dispersions relations. Particles have a minimum
group velocity that is $1/3$ of the speed of light, so that there are
no particle excitations at rest, even though there is a mass gap.
This phenomenon is due to the collective excitations.
Holes have the same group velocity for $\kk = 0$ but with
an opposite sign, and a minimum energy for approximately
$\kk=0.408\,m_f$\ref{4}. All these features combined together
make the thresholds difficult to determine.
The dispersion relations for particles and holes
are displayed schematically in Fig.\ 1.

Before studying the thresholds
of the diagram in Fig.\ 2 we need an
expression for $\Im\Sigma$ including all possible
channels. We may use
the expression in\ref{3} suitably modified to account
for collective excitations and chirality. One difference between the
right and left handed fermion propagators is  thermally induced
masses $m_{L,R}$ which are not equal because the right and left handed fermions
are  in different representations of the Standard Model
gauge group. The one--loop resummed propagator in the rest frame of the heat
bath
reads $(P_\mu=(p_0,\vek p)$ with $\pp =|\vek p|)$
$$
      S(P)=\eta\frac{L\slask(P)}{L^2(P)}\bar{\eta}
      +\bar{\eta}\frac{R\slask(P)}{R^2(P)}\eta\ ,
\eqno\new)
$$
where\eqnam{\fourth}
$$\eqalign{
      \eta&=\inv{2}(1-\gamma_5)\ ,\ \
      \bar{\eta}=\inv{2}(1+\gamma_5)\ ,\cr
      L\slask&=(1+a_L(p_0,\pp))P_\mu\gamma^\mu+
      b_L(p_0,\pp)\gamma^0,\cr
  a_L(p_0,\pp)&=\frac{m_L^2}{\pp^2}\left(
      1-\frac{p_0}{2\pp}\ln\Abs{\frac{p_0+\pp}{p_0-\pp}}\right),\cr
       b_L(p_0,\pp)&=\frac{m_L^2}{\pp}\left(
      -\frac{p_0}{\pp}+(\frac{p_0^2}{\pp^2}-1)
      \inv{2}\ln\Abs{\frac{p_0+\pp}{p_0-\pp}}\right)\ ,\cr
}
\eqno\new)
$$
and similarly for $R\slask$.
The functions $a_{L,R}(p_0,\pp)$ and
$b_{L,R}(p_0,\pp)$ are respectively even and
odd with respect $p_0$, i.e.\ they have the properties
\eqnam{\abprop}
$$
      a_{L,R}(-p_0,\pp)=a_{L,R}(p_0,\pp)\ ,\ \
      b_{L,R}(-p_0,\pp)=-b_{L,R}(p_0,\pp)\ .
\eqno\new)
$$
We have introduced left and right handed  fermion
masses in \eqt{\fourth}. For leptons one finds
$$\eqalign{&\cr
m^2_{l,L} &= {m_Z^2 + 2m_W^2 +  m_l^2 + m_{l'}^2\over 2 f^2}T^2,\cr
m^2_{l,R} &= {m_Z^2 - m_W^2 + \frac 12 m_l^2\over 2 f^2}T^2,\cr&\cr
}\eqno\new)
$$
where $m_l$ and $m_{l'}$ are the $T=0$ masses of the leptons $l$ and $l'$
belonging to the same $SU(2)$ doublet. For quarks one finds that
$$\eqalign{&\cr
m^2_{u_i,L} &= \frac 16 g_s^2 T^2 + { 3 m_W^2 + \frac 19 (m_Z^2 - m_W^2) +
                m^2_{u_i} + m^2_{d_i}\over 8 f^2}T^2,\cr
m^2_{u_i,R} &= \frac 16 g_s^2 T^2 + {\frac 49 (m_Z^2 - m_W^2) + \frac 12
               m_{u_i}^2\over 2f^2}T^2,\cr&\cr
}\eqno\new)
$$
and
$$\eqalign{&\cr
m^2_{d_i,L} &= \frac 16 g_s^2 T^2 + { 3 m_W^2 + \frac 19 (m_Z^2 - m_W^2) +
                m^2_{u_i} + m^2_{d_i}\over 8 f^2}T^2,\cr
m^2_{d_i,R} &= \frac 16 g_s^2 T^2 + {\frac 19 (m_Z^2 - m_W^2) + \frac 12
               m_{d_i}^2\over 2 f^2}T^2,\cr&\cr
}\eqno\new)
$$
where $ m_{u_i}$ and $m_{d_i}$ are the $T=0$ masses of $u$
and $d$ type quarks of
$i$-th generation, respectively, $m_W=80.0\ {\GeV},\ m_Z=91.2\ \GeV$
and $m_t$ is the $T=0$ mass of the top-quark.
The higgs vacuum expectation value is $f\simeq 247$ GeV. For the
strong coupling constant we adopt the value $g_s \simeq 1.2$.
Note that the left handed masses of fermions
belonging to same $SU(2)$ doublet are equal due to unbroken symmetry.

The imaginary part of the self-energy of the diagram in Fig.\ 2
is
\eqnam{\fermiImS}
$$
\eqalign{
      \Im\Sigma_f(p_0,\pp)=&
      -\frac{4g_Y^2}{\sinh 2\theta_p}\int\frac{d\,^4K}{(2\pi)^2}
      \epsilon(k_0)\epsilon(k_0-p_0)\inv{4}\sin 2\phi_k
      \sin 2\phi_{k-p} \cr
      &\times L_\mu(K)R^\mu(K-P)\eta_L\eta_R
      \delta(L^2(K))\delta(R^2(K-P))\ ,
\cr}
\eqno\new)
$$
where
$$
      \sin 2\phi_k=2\frac{e^{\beta|k_0|/2}}{e^{\beta|k_0|}+1}\ ,
\eqno\new)
$$
and $g_Y$ is the corresponding Yukawa coupling.
The factors $\eta_L$ and $\eta_R$ are $+1$ and $-1$ for particles
and holes, respectively. They arise when the energies in
$L^2(K)$ and $R^2(K-P)$ are continued analytically using the time--ordered
prescription
$k_0\rightarrow k_0+i\epsilon k_0$. For a free massive
propagator this would only add $i\epsilon$ in the denominator.
In our case we have
$$
      L^2(k_0+i\epsilon k_0,\vek{k})=L^2(K)+
      i\epsilon\, k_0{\partial L^2\over \partial k_0}\, .
\eqno\new)
$$
The factor  that multiplies $i\epsilon$ is positive for particle solutions
and negative for hole solutions which gives rise to an
extra sign for the hole contribution. Below we shall see that this extra sign
is crucial in order to get a positive damping rate in
all channels, whereas the naive propagator $(L^2(K)+i\epsilon)^{-1}$
would lead to an instability.

\centerline{\subsection 2.2 Thresholds}
\jump
{}From the analysis in\ref{15} it follows that there are two possible decay
channels for a scalar interacting with
fermions with two different masses,
$m_1$ and $m_2$.
In the case of a scalar at rest $(p_0 = m_H, \vek{p}=0)$
they are: (1) decay if
$m_H>m_1+m_2$, (2) absorption if
$m_H<|m_1-m_2|$.
The inverses of these processes are also possible and they
are included in the rates we compute below. Here, and in what follows, we
denote by $m_H\equiv m_H(T)$ the $T$--dependent higgs mass.

In our case the thresholds
are not determined by the $\vek{k}=0$ energy but must be derived using the
complete dispersion relations. A decay
 or absorption can take place when one of the following
conditions is fulfilled
\eqnam{\kincond}
$$
\eqalign{
      (1)&\ \ m_H=\omega_{L,i}(\kk)+\omega_{R,j}(\kk)\ ,\cr
      (2)&\ \ m_H+\omega_{L,i}(\kk)=\omega_{R,j}(\kk)\ ,\cr
      (3)&\ \ m_H+\omega_{R,j}(\kk)=\omega_{L,i}(\kk)\ .
\cr}
\eqno\new)
$$
Here k$=|\vek k|$ and for each energy we have put an index ($i,j=p,h$) to
distinguish
between particles and holes. However, not all combinations of
particles and holes contribute in each channel because
the factor $L_\mu R^\mu$ can vanish. This turns out to
be important in order to obtain a positive thermalization
rate.

Let us denote by
$E$ the energy of an excitation,
including a sign with reference to the arrows
in Fig.\ 2. Then $E_L$ and $E_R$ have opposite signs in the decay process, and
equal signs in the absorption process. From \eqt{\fourth} we have explicitly
\eqnam{\LR}
$$
\eqalign{ &\cr
      &L_\mu(E_{L,i},\kk) R^\mu(E_{R,j},\kk)= \cr
      &[E_{L,i}\bigl(1+a_L(E_{L,i},\kk)\bigr)
            +b_L(E_{L,i},\kk)]
      [E_{R,j}\bigl(1+a_R(E_{R,j},\kk)\bigr)
            +b_R(E_{R,j},\kk)]\cr
      &-\kk^2 \bigl(1+a_L(E_{L,i},\kk)\bigr)
            \bigl(1+a_R(E_{R,j},\kk)\bigr)\ .
\cr &\cr}
\eqno\new)
$$
This should be compared with the on-shell conditions for
positive energy particles and holes
\eqnam{\masscond}
$$
\eqalign{
      \omega_p(1+a(\omega_p,\kk))+b(\omega_p,\kk)
      &=\kk(1+a(\omega_p,\kk))\ ,\cr
      \omega_h(1+a(\omega_h,\kk))+b(\omega_h,\kk)
      &=-\kk(1+a(\omega_h,\kk))\ .
\cr}
\eqno\new)
$$
Negative energy solutions, where $\omega\rightarrow-\omega$
in \eqt{\masscond}, are also possible but
then the on-shell conditions for particles and holes are
reversed due to the properties of $a_{L,R}$ and
$b_{L,R}$ in \eqt{\abprop}.
Using \eqt{\masscond} in \eqt{\LR} we see that when
$E_L$ and $E_R$ have the same sign, only ($i=p,j=h$)
and ($i=h,j=p$) give a non-zero result. When
$E_L$ and $E_R$ have opposite signs we are left with ($i=p,j=p$)
and ($i=h,j=h$). This selection rule
can be understood  physically as a consequence of
the conservation of angular momentum. The
chirality--helicity relation is reversed
for holes\ref{5}, and because the fermions in scattering
processes under consideration always have different
chiralities, absorption is possible only if they
have same helicities, and direct decay
is possible only if they have opposite helicities.

For a Higgs particle at rest  the angular
integrals in \eqt{\fermiImS} become trivial and the $\delta$-functions
can be used to perform the $k_0$ and $\kk$ integrations. One
should note that the $\delta$-function may be rewritten as
$$
      \delta(L^2) =
      \sum_{i=p,h} W_{L,i}(\kk)[\delta(k_0 - \omega_{L,i}(\kk)) +
      \delta(k_0 + \omega_{L,i}(\kk))]\ ,
\eqno\new)
$$
where the energy factor $W_{L,i}$ is different from the usual $1/2\omega(\kk)$.
In fact, we have instead
$$
      W^{-1}_{L,i}(\kk)=
      \Abs{\frac{\del L^2}{\del k_0}}_{k_0=\omega_{L,i}}\ ,
\eqno\new)
$$
and similarly for $W_R$.
Combining the expression in \eqt{\LR} for $L_\mu R^\mu$ with the factors
$W_L$ and $W_R$ we find the rather simple expression
\eqnam{\lrww}
$$
      L_\mu R^\mu W_L W_R =
      -\frac{2(\omega^2_{L,i}-\kk^2)(\omega^2_{R,j}-\kk^2)}
      {4m^2_L\,4m^2_R}\ ,
\eqno\new)
$$
which will be useful later. We note that $L_\mu R^\mu W_L W_R$
is always negative.

\jump
\centerline{\subsection 2.3 Decay rates}
\jump

Let us start with the absorption channel, which  opens up
when $m_H$ is small enough. Its existence is determined
by the last two conditions in \eqt{\kincond}.
In this case $E_L$ and $E_R$ have the same sign so that
only $p\to h$ and $h\to p$ processes contribute.
By virtue of the dispersion relations (see Fig.\ 1)
there is a channel for small $m_H$ even
if $m_L=m_R$, because when $\omega_h\leq\omega_p$
the higgs may be absorbed by a
hole in the heat bath to produce a particle.
Using \eqt{\lrww} the imaginary part for the process $H+f_{R,h}\rightarrow
f_{L,p}$
can then be found to be
\eqnam{\HRhLp}
$$
      \Im\Sigma (Rh,Lp) =\frac{4g_Y^2}{\pi}\eta_{L,p}\eta_{R,h}\kk^2
      \frac{(\omega^2_{L,p}-\kk^2)(\omega^2_{R,h}-\kk^2)}
      {4m^2_L\,4m^2_R}
      \bigl[n_{R,h}(\kk)-n_{L,p}(\kk)\bigr]\ ,
\eqno\new)
$$
where k is determined by the relation
$$
      m_H+\omega_{R,h}(\kk)=\omega_{L,p}(\kk)\ ,
\eqno\new)
$$
and
$$
      n_{L,R}= \frac{1}{e^{\beta\omega_{L,R}}+1}\ ,
\eqno\new)
$$
are the fermion distribution functions.
We find that $\Im\Sigma$ in \eqt{\HRhLp} is negative, as it should, leading to
a positive decay rate, but the reason is different from the case
of ordinary massive fermions\ref{15}. Here, the matrix element
$L_\mu R^\mu$ is negative in all channels, in contrast
to the case in\ref{15} where the signs are different for
decay and absorption. On the other hand, we have an extra factor
of $\eta_{L,i}\eta_{R,j}$ which then helps to
give the correct sign. We note that this
works because only $p\to h$ and $h\to p$ processes contribute in
this channel. To obtain the complete fermionic $\Im\Sigma$ for absorptions,
the contribution from the process
$H+f_{L,h}\rightarrow f_{R,p}$ should be added
to \eqt{\HRhLp} so that $\Im\Sigma_{abs}=\Im\Sigma (Rh,Lp)+\Im\Sigma (Lh,Rp)$.
In
the minimal Standard Model the difference between $m_L$ and
$m_R$ is so small compared to the particle--hole difference
that we could safely approximate $m_L=m_R$, which would simplify the expression
for $\Im\Sigma_{abs}$. (See, however, the discussion
below when the mass difference is not neglected.)

{}From Fig.\ 1 we see that the threshold is
at a finite $\kk_1$ and two solutions around $\kk=\kk_1$
exist for $m_H$ slightly below the threshold. When $m_H$
decreases one solution goes to zero, if we neglect the
mass difference, and therefore its contribution to
$\Im\Sigma$ goes to zero. The other solution goes to
infinity but the residue of the hole propagator, which
is proportional to $\omega^2_h-\kk^2$, goes exponentially
to zero and thus $\Im\Sigma{abs}\rightarrow 0$ when
$m_H\rightarrow 0$.

If we do not neglect the mass difference in the absorption
channel the threshold structure becomes more complicated. Let us
assume that $m_L>m_R$. Then in the $H+f_{R,h}\rightarrow f_{L,p}$
channel (Fig.\ 1a) one of the solutions disappears when $m_H<m_L-m_R$.
In the $H+f_{L,h}\rightarrow f_{R,p}$ channel (Fig. 1b) the dispersion
curves cross at some point $\kk_2$.
The threshold is determined by
the largest difference to the right of the crossing point
($\kk_1$ in Fig.\ 1b).
As $m_H$ decreases
there appear two solutions to the right.
In the limit $m_H\rightarrow 0$, the large k solution
disappears since the hole residue goes exponentially to zero.
We are then left with a solution which for a small mass difference
is located at $\kk_2=\frac{3}{2}(m_L-m_R)$.
When $m_H\rightarrow 0$ the thermalization rate can be found
from \eqt{\HRhLp} to be
\eqnam{\gF}
$$
      \gamma_F/T=
      {9g_Y^2\over16 \pi}\left ( {(m_R+m_L)^2\over 4m_Rm_L}\right )^2
       \left ({m_R - m_L\over T}\right ) ^2
      {1\over 1 + \cosh[\beta (m_R+m_L)]}\ .
\eqno\new)
$$

There is also a third absorption channel, $H+f_{R,p}\rightarrow f_{L,h}$,
which opens for $m_H<m_L-m_R$ and k is located between
0 and $\kk_2$. The thermalization rate in the limit $m_H\rightarrow 0$
is again given by \eqt{\gF}.

The direct decay channel is  simpler. The signs of $E_L$ and $E_R$
are opposite, and the higgs decays either
into a pair of particles or into a pair of holes. Therefore, the
product $\eta_L\eta_R$ is +1, leading again to a positive
decay rate. Summing over both decay channels we find
\eqnam{\fdecay}
$$
      \Im\Sigma_{dec}=-\frac{4g^2}{\pi}\sum_{i=p,h}\kk^2
      \frac{(\omega^2_{L,i}-\kk^2)(\omega^2_{R,i}-\kk^2)}
      {4m^2_L\,4m^2_R}(1-n_{L,i}-n_{R,i})\ ,
\eqno\new)
$$
and k is in this case determined by
$$
      m_H=\omega_{L,i}(\kk)+\omega_{R,i}(\kk)\ .
\eqno\new)
$$
The ordinary $T=0$ contribution is included in this expression.
The threshold for holes is lower than for particles
since the minimum energy for holes is about $0.928\,m_f$\ref{4}.
There are also two solutions in this channel when
$1.865 \, m_f< m_H < 2\, m_f$. As $m_H$ increases the corresponding k
increases and the hole residue goes exponentially to zero so
that the particle channel dominates as soon as $m_H$ is above
the particle threshold.

The full fermionic contribution is thus finally given by
$\Im\Sigma_f=\Im\Sigma_{abs}+\Im\Sigma_{dec}$. In practise the direct decay
channel is in many cases irrelevant, because its existence would imply
that $m_H(T)\gsim T$. At least in perturbation theory $m_H$ is expected to be
less than $T$ in the high temperature limit. We shall return to $\Im\Sigma_f$
in Section 4, where
we also plot it for the top quark.

\jump
\centerline {\section 3. Gauge bosons}
\jump
\centerline {\subsection 3.1 One-loop case}
\jump

At $T=0$ the imaginary part of the diagram of Fig.\ 3 is zero for an
on-shell massive
Higgs particle and a massless gauge boson for kinematical
reasons. At finite temperature  thermal
masses are induced for both the Higgs and the gauge boson so that an
absorption
channel opens when $2\,m_H(T) < m_W(T)$. In particular, this condition may be
fulfilled close
to the critical temperature of a second order, or a weakly
first order, phase transition where
the higgs mass becomes small.

The self-energy of a gauge boson is a rather complicated function of
the momentum\reftwo{1}{2}. We may, however, still use the standard
expression for the imaginary part of the self-energy, which
can be found in e.g.\ref{3}, if we take into account the
fact that the higgs couples to gauge bosons through a derivative
coupling, and replace the $\delta$-functions with the ones
containing the thermal on-shell condition.
With the high $T$
expression for the gauge boson propagator\footnote{*}{\bx We are using the
Feynman gauge throughout the paper.}\reftwo{1}{2}
we get ($K_\mu=(k_0,\vek{k}),\kk=|\vek{k}|$)
\eqnam{\gbImS}
$$
\eqalign{
      &\Im\Sigma_{gb}(p_0,\vek{p})
      = \cr
      &-\frac{e^2}{\sinh 2\theta_p}
      \int\frac{d\,^4K}{(2\pi)^2}\inv{4}\sinh 2\theta_k
      \sinh 2\theta_{p-k}\; \delta((\vek{p}-\vek{k})^2-m_H^2(T))\cr
      &\times\Big\{ 4\bigg(
      \frac{(\vek{p}\cdot\vek{k})^2}{\kk^2}-\vek{p}^2\bigg)
      \delta(K^2-\pi_T)+\frac{4}{K^2\kk^2}
      (p_0\kk^2-k_0\vek{p}\cdot\vek{k})^2
      \delta(K^2-\pi_L)\Big\}\ , \cr
      }
\eqno\new)
$$
where
$$
      \sinh 2\theta_p =
      2\frac{e^{\beta|p_0|/2}}{e^{\beta|p_0|}-1}\ .
\eqno\new)
$$
and
$$
e^2=\cases{&$\frac 34 g^2\ \ {\rm for\ }SU(2)$\ ,\cr&\cr
          &$g'^2\ \ {\rm for\ }U(1)$\ .\cr} \eqno\new)
$$
The gauge boson self-energy has a transverse and a
longitudinal part, which are given by
$$\eqalign{&\cr
      \pi_T(k_0,\vek{k})&=\frac{g^2T^2}{6}
      \left((n_W+1)\left[\frac{k_0^2}{\kk^2}
      +(1-\frac{k_0^2}{\kk^2})\frac{k_0}{2\kk}
      \ln\Abs{\frac{k_0+\kk}{k_0-\kk}}\right]-2\right)\ ,\cr \cr
       \pi_L(k_0,\vek{k})&=\frac{(n_W+1)g^2T^2}{3}
      (1-\frac{k_0^2}{\kk^2})
      \left[1-\frac{k_0}{2\kk}
      \ln\Abs{\frac{k_0+\kk}{k_0-\kk}}\right]\ .\cr&\cr}
\eqno\new)
$$
These self-energies include the contributions from gauge
boson, fermion and  scalar
loops. We use the notation
$n_W=(N_f+4N)/4$, where $N_f=2\times(\#\  of\ generations)$
and $N=2$ for $SU(2)$ gauge fields. For $U(1)$
gauge field $g$ is replaced by $g'$
and $n_W$ by $n_Y = {19\over 36} N_f$. We define also
the longitudinal and transverse gauge boson masses
by $M_L^2 = \pi_L(\omega,\vek k = 0)$ and
$M_T^2 = \pi_T(\omega,\vek k = 0)$, respectively.

At one loop the high $T$ expression for the
higgs mass in the unbroken phase is given by
$$
      m^2_H(T)=-\frac{m^2_H(0)}{2}
      +\frac{6\,m^2_W+3m^2_Z+6\,m^2_t + m^2_H(0)}{12f^2}\,T^2\ ,
\eqno\new)
$$
where $m_H(0)$ is the zero temperature higgs mass.
The longitudinal masses of $SU(2)$ and $U(1)$ gauge bosons are given by
$$\eqalign{ &\cr
     M^2_{L,SU(2)}&={n_W + 1\over 9}\frac{4m^2_W}{f^2}T^2,   \cr
              M^2_{L,U(1)}&={n_Y + 1\over 9}\frac{4(m^2_Z-m^2_W)}{f^2}T^2,\cr
&\cr
}
\eqno\new)
$$
while the transverse masses are
$$\eqalign{ &\cr
     M^2_{T,SU(2)}&= {n_W - 2\over 9}\frac{4m^2_W}{f^2}T^2,\cr
    M^2_{T,U(1)}&= {n_Y - 2\over 9}\frac{4(m^2_Z-m^2_W)}{f^2}T^2.\cr &\cr
}
\eqno\new)
$$
It is the scalar loop that gives rise to the difference between
the longitudinal and transverse masses.

For a Higgs particle at rest ($p_0=m_H, \vek{p}=0$)
only the longitudinal part contributes and the angular
integral becomes  again trivial. We obtain, as in
Section 2, an energy factor   from the residue at the pole, given by
$$
      \widetilde{W}^{-1}_L(\kk)=
      \Abs{\frac{\del}{\del k_0}
      \Bigl(K^2-\pi_L(k_0,\kk)\Bigr)}_{k_0=\omega_L(\kk)}
      =\frac{M_L^2-\pi_L(\omega_L(\kk),\kk)}{\omega_L(\kk)}\ .
\eqno\new)
$$
The absorption channel opens up when the higgs mass is so small that
it can annihilate with a higgs from the heat bath to produce
a gauge boson. In that channel $\Im\Sigma$ is given by
$$\eqnam{\gbim}
      \Im\Sigma_{gb}=-\frac{e^2}{\pi}\frac{\widetilde{W}_L(\kk)}
      {\pi_L(\kk)\sqrt{\kk^2+m_H^2(T)}}
      m_H^2(T)\kk^4\bigl[n_H(\kk)-n_L(\kk)\bigr]\ ,
\eqno\new)
$$
where $\kk$ is determined by solving the on-shell condition
of the gauge boson relation\eqnam{\gbrelation}
$$
      \omega_L^2(\kk)=\kk^2+\pi_L(\omega_L(\kk),\kk)\ ,
\eqno\new)
$$
and the energy conservation condition\eqnam{\econservation}
$$
      m_H(T)+\sqrt{\kk^2+m_H^2(T)}=
      \sqrt{\kk^2+\pi_L(\omega_L(\kk),\kk)}\ .
\eqno\new)
$$
The thermal distribution functions are
$$
      n(k)_{H,L}=\inv{e^{\beta\omega_{H,L}}-1}\ ,
\eqno\new)
$$
with
$$
      \omega_H=\sqrt{\kk^2+m_H^2(T)}\ ,\ \
      \omega_L=\sqrt{\kk^2+\pi_L}\ .
\eqno\new)
$$
Equation (\gbim) expresses the imaginary part for any
component of the Higgs doublet;
they are all equal because of the unbroken gauge symmetry.

We are mainly interested in the thermalization rate
close to the phase transition where $m_H(T)$ is
expected to be small. In the limit
$m_H \rightarrow 0$ the thermalization rate due to $SU(2)$ gauge bosons is
given by\eqnam{\iirokaava}
\def\ee{(n_W+1)g^2}
$$\eqalign{
      \gamma_{gb}/T &=
      {3g^2\over 8\pi}\left ( {T\over M_L}\right ) ^2
      \left ({\ee\over 6}  \ln{T\over
      m_H}\right ) ^{3/2}\cr
      &\qquad {m_H\over T} \exp \left [ -\left ({\ee\over 6} \ln{T\over m_H}
      \right )^{1/2} \right ].\cr} \eqno\new)
$$
This result can be found by using the asymptotic formula\ref{1}
$\pi_L=4\kk^2 \exp\left (-{6\kk^2/(n_W+1)g^2T^2}\right )$ which
is valid for $3\kk/\sqrt{n_W+1}\,gT\gg 1$, together with Eqs. (\gbrelation )
and (\econservation ). One solves  $\kk$ iteratively as a function of $m_H$,
whence
one obtains \eqt{\iirokaava} from \eqt{\gbim}, and the expansion is valid when
$m_H\lsim
10^{-3}T$, corresponding to $\gamma_{gb}/T\lsim 1.9\times 10^{-4}$.

\jump
\centerline {\subsection 3.2 Two-loop case}
\jump

At two--loop level the variety of relevant processes is much larger
than at one loop,
leading to tens of different diagrams. We now estimate the order
of magnitude of two--loop processes using the two
representative graphs given in Fig.\ 4. The purely scalar "rising sun" diagram
of
Fig.\ 4a has been calculated in\ref{20}. The imaginary part
of the two--point function for an on--shell complex
higgs with a vanishing external three momentum is, in the leading high $T$
approximation,
\eqnam{\gHtwo}
$$
{\rm Im}\Sigma_H = -{\lambda^2\over 256\pi}T^2\, .
\eqno\new)
$$
We may use this result to estimate the imaginary part of Fig.\ 4b which then
reads
$$ \eqnam{\gbrs}
{\rm Im} \Sigma = - {3g^4\over 256\pi}T^2
\eqno\new)
$$
for $SU(2)$ gauge bosons. The $U(1)$ gauge boson has a similar expression but
with
$3g^4$ replaced by $4 (g')^4$. In the derivation
of \eqt{\gbrs}
the mass difference of gauge bosons and Higgs bosons was neglected;
formally, the omitted terms are next--to--leading corrections in
terms of the gauge couplings. The two--loop thermalization rate
is thus\eqnam{\twoloop}
$$
\gamma^{(2)}_{gb}/T \simeq {3g^4\over 256\pi}{T\over m_H}\ .
\eqno\new)
$$
When $m_H$ decreases  $\gamma_{gb}$ increases. Therefore, for very
small higgs masses the two--loop processes could actually dominate
over the one--loop ones, for which the rate goes to zero as $m_H\to 0$ as
is evident from \eqt{\iirokaava}.

There are, however, certain issues which make our two--loop
estimate \eqt{\twoloop} uncertain. First, we expect $\Im\Sigma(p_\mu=0)$
to be zero (which it formally is in the full
two--loop expression in\ref{20}) but infrared dominance
leads to the constant in \eqt{\gHtwo}, which then gives a
diverging $\gamma_H$ when $m_H\rightarrow 0$.
Secondly, the dispersion relation with a
complex $\Sigma$ reads
\eqnam{\Cdp}
$$\eqalign{
      \omega^2 &= \pp^2+m^2_H(T)+\frac{\gamma^2}{4},\cr
      \gamma &= -\frac{\Im\Sigma}{\omega}\ ,\cr}
\eqno\new)
$$
which modifies the simple relation
$\omega^2=\pp^2+m^2_H(T)$ when $m^2_H(T)\lsim\Abs{\Im\Sigma}$.
(Note that $\Re\Sigma$ has been included in the definition of the
renormalized mass $m_H$.)
Also, when propagators have complex poles the
$\delta$-functions in \eqand{\fermiImS}{\gbImS} should be replaced by Lorentz
functions.
Taken all together, we conclude that the simple comparison with the ``rising
sun"
diagram gives a indication that the two--loop contributions may be
comparable to the one--loop ones, although the actual magnitude remains
somewhat
uncertain.

\jump
\centerline {\section 4. Thermalization rate vs.\
      phase transition time}
\jump

As an application, let us now turn to consider a first--order
electroweak phase transition. There has been a number of attempts\ref{21}
to
derive the form of the appropriate effective potential for the
Higgs field in perturbation theory, but lattice studies\ref{22} seem to
indicate that there also exists a large non--perturbative component.
Therefore, to be definite, we shall merely assume that
the free energy of the neutral Higgs field for $\phi\simeq 0$ and
near the critical
temperature $T_c$ can be expressed through the expansion \eqnam{\pot}
$$
V(\phi)=\frac 12 \kappa (T^2-T^2_0)\phi^2-\frac 13\alpha T\phi^3
+\frac 14\lambda\phi^4+\dots .
\eqno\new)
$$
where $\kappa$, $\alpha$, $\lambda$ and $T_0$ are (unknown) parameters
with $T_c^2=T^2_0/(1-2\alpha^2/9\lambda\kappa)$. In principle, they need not
be equal to the corresponding parameters at $T\simeq 0$.
Thermodynamical properties of \eqt{\pot}, as well as bubble nucleation, has
been discussed in\ref{25}. Nucleation starts at $T_f$ where
$T_0<T_f<T_c$, and the time scale after which the major part of a
given volume is filled by bubbles of new phase can be estimated to
be \eqnam{\growth}
$$
      t_{\rm growth}=-{1\over S'(t_f)}=t_c
      {\frac 12 A^{\frac 12}\over \ln^{\frac 32}(M^4_P/T_c^4)
}\ ,
\eqno\new)
$$
where $S(t)$ is the action, and in the small supercooling limit
$A=0.0144\,\alpha^5\lambda^{-7/2}\kappa^{-2}(T_c/T_0)^4$. Here
$t_c$ is the time corresponding to $T_c$ and is related
to the Hubble rate by $H=(8\pi^3g_*/90)^{1/2}T_c^2/M_P=1/2t_c$.
The size of the critical bubble, $R_c$, depends on the parameters of
the potential \eqt{\pot} but is typically larger than the correlation
length\ref{23}.

Thermal fluctuations generate configurations which may
locally affect the dynamics of the phase
transition, and for that purpose it is important to find out
the lifetime of these  fluctuations. In particular, fluctuations larger
than the correlation length, if sufficiently stable, could be important
for bubble formation and growth.

In the previous Sections we have computed the thermalization rate
for perturbations with small amplitude ($b(x)^2\le
(p^2+m^2_H)/\lambda$) and  large spatial size (${\bf p}\simeq 0$);
these correspond then to large subcritical "bubbles".
Since the precise $T$--dependence of $m_H$ close to $T_c$ is not reliably
described by the loop expansion, we prefer to parametrize $\gamma$ as a
function of $m_H$ rather than $T$.
When varying $m_H/T$
we can consider $T$ to be fixed and vary the zero temperature higgs mass
$m_H(0)$, which is not experimentally known. If we rather want $\gamma$ as
a function of $T$, we would have to take into account the $T$--dependence of
$m_H$, and the fact that the vertical scale in Fig.\ 5 depends on $T$.

The largest
fermionic contribution to the imaginary part comes from the top
quark intermediate states, and in Fig.\ 5 we have drawn
the fermionic thermalization rate, assuming $m_t=135$ GeV and $m_L=m_R$,
together with the gauge
boson contributions.
The thresholds are such that the one--loop rate
is zero when $0.25T\lsim m_H(T)
\lsim 1.15T$. (The treshold structure for the bottom quark is different from
that
of the top quark and would actually yield a non--zero thermalization rate
in  parts of the region  $0.25T\lsim m_H(T)
\lsim 1.15T$, but with a relatively small rate.) In contrast, the two--loop
rate does not have any thresholds so
that thermalization will in fact take place for all values of $m_H$. Comparing
with Fig.\ 5, we find that the
two--loop estimate \eqt{\twoloop} would become bigger than the one--loop
rate when $m_H/T\lsim 5\times 10^{-2}$.

{}From Fig.\ 5
we see also that the thermalization rate is roughly
$\gamma\sim 10^{-2}T\gg H$ in the region
where $m_H$ is small. Thus compared with the Hubble time,
thermalization is very fast. However, the phase transition may
also be very fast. This is so
especially when the amount of supercooling is small.
In that case typically $m_H\ll T$ during the transition
and we may safely approximate $t_f\simeq t_c$.
 Comparing thermalization rate with the growth time we find that
\eqnam{\compare}
$$
\gamma\;t_{\rm growth}\simeq 7\times 10^{-8}A^{1\over 2}\left ({M_P\over
T_c}\right) .
\eqno\new)
$$
At $T_c$ we may take for reference values the one--loop results $\kappa=0.23$
(with
$m_t=135\; \GeV)$ and $\alpha=0.028$, and for $\lambda$ an arbitrary value
0.006
to satisfy the baryon--number retention bound\ref{24} (in
$\lambda\phi^4$--theories there are actually
indications\ref{26} that $\lambda$ could be driven to very
small values at $T_c$). For these values $A=0.35$. With $T_c\simeq 100$ GeV, we
then find that $\gamma\;t_{\rm growth}\simeq 5\times 10^9$.
In order that $\gamma\;t_{\rm growth}\sim 1$ we would need $A\sim 10^{-20}$,
or extremely small $\alpha$ and phase transition which is almost second
order. Although in principle this might be possible,
we conclude that it is unlikely that spatially large fluctuations can
remain stable during the electroweak phase transition.
\jump

\centerline {\section 5. Summary and conclusions}
\jump
We have calculated the one--loop thermalization rate of the
Higgs field in the high temperature phase in the Standard Model.
A fluctuation of the Higgs field thermalizes through
scattering processes with itself,
with the gauge bosons and with the fermions.
It is necessary to resum the hard thermal loops
for the gauge boson and the fermion self--energies in order to get
any kinematically allowed processes at the one--loop level. This changes
the dispersion relations and induces thermal masses in such a way
that the Higgs particle can annihilate with a higgs from the heat bath to
create a gauge boson, be absorbed by a fermion,
or decay into a pair of fermions. The inverses of these processes
are also possible.
By using resummed propagators for gauge bosons and fermions
we consider the external states to be the effective particle
excitations in the heat bath, including collective excitations.

It is the existence of the two fermionic branches at high $T$,
particles and holes with opposite chirality--helicity relations, that allow for
a Higgs particle to be absorbed
by fermions in the heat bath. The unusual dispersion relation
for holes makes the determination of energy thresholds
complicated and it has to be done numerically.
To get a positive thermalization rate one has to
use propagators with the correct analytical
continuation, $k_0\rightarrow k_0+i\epsilon k_0$, and not just
add $i\epsilon$ in  the denominator.  Only
processes which conserve helicity
are possible, and this is also essential in order to
get a positive thermalization rate.

Two Higgs particles can annihilate and produce a gauge boson
when the higgs mass is small enough. There are, of course,
also higher order processes that do not have any thresholds.
In order to make a crude estimate of the two-loop processes we
estimate them by a two-loop
gauge boson  "rising sun" diagram.
We found that for the Standard Model parameters there is a possibility that
the two-loop effects
are of the same order of magnitude as the one-loop ones.

The thermalization rate is compared with other time
scales in Section 4. Two such scales of interest are the Hubble expansion and
the electroweak phase transition rates. The time it takes the bubbles of new
phase to fill up the universe is to
a large extent model dependent but can be estimated in
a phenomenological model\ref{25}. We found that in general the
thermalization rate is several orders of magnitude faster than
both the Hubble rate and the phase transition rate. This
does not mean that thermal fluctuations are unimportant
for the dynamics of the phase transition. Rather, it justifies
time averaging over fluctuations when comparing them with the
phase transition rate.

\vfill \eject\null
\centerline {\subsection References}
\jump

\refitem{1}
      {H. A. Weldon,
      {\sl ``Covariant calculations at finite temperature:
      The relativistic plasma"},
      \pr{D26}{1982}{1394}.}
\refitem{2}
      {V.~V.~Klimov,
      {\sl ``Collective excitations in a hot quark-gluon
      plasma"},
      {\it Sov. Phys. JETP} {\bf 55} (1982) 199.}
\refitem{3}
       {R.~L.~Kobes and G.~W.~Semenoff,
      ``{\sl Discontinuities of Green functions in field theory
      at finite temperature and density~}",
      \np{260}{1985}{714}.}
\refitem{4}
     {R. D. Pisarski,
    {\sl ``Renormalized fermion propagators in hot gauge theories"},
      \np{A498}{1989}{423c}.}
\refitem{5}
      {H. A. Weldon, ``{\sl Dynamical holes in the quark-gluon
       plasma"},
      \pr{D40}{1989}{2410}.}
\refitem{6}
      {J.~Iliopoulos, C.~Itzykson and A.~Martin,
      {\sl ``Functional methods and perturbation theory"},
      \rmp{47}{1975}{165.}}

\refitem{7}
      {R.~D.~Pisarski,
      {\sl ``Computing finite-temperature loops with ease"},
      \np{B309}{1988}{476.}}

\refitem{8}
      {A.~Weldon,
      {\sl ``Mishaps with Feynman parametrization
      at finite temperature"},
      \pr{D47}{1993}{594.}}

\refitem{9}
        {R.~R.~Parwani,
        {\sl ``Resummation in a hot scalar field theory"},
        \pr{D45}{1992}{4695.}}

\refitem{15}
      {A.~Weldon, ``{\sl Simple rules for discontinuities
      in finite-temperature field theory~}",
      \pr{D28}{1983}{2007.}}

\refitem{16}
      {L.~Dolan and R.~Jackiw,
      ``{\sl Symmetry behavior at finite temperature~}",
      \pr{D9}{1974}{3320.}}

\refitem{17}
      {S.~Coleman and E.~Weinberg,
      {\sl ``Radiative corrections as the origin of spontaneous
      symmetry breaking"}
      \pr{D7}{1973}{1888.}}

\refitem{20}
       {P. Elmfors, K. Enqvist and I. Vilja,
      {\sl ``Finite temperature effective action and
      thermalization of perturbations"}, preprint NORDITA - 93/22 P.}

\refitem{21}
      { For a recent review see A.~G.~Cohen, D.~B.~Kaplan and A.~E.~Nelson,
      {\sl ``Progress in electroweak baryogenesis"},
      to appear in {\it Annual Review of Nuclear and
      Particle Science } {\bf 43}, and references therein.}

\refitem{22}
      {K.~Kajantie, K.~Rummukainen and M.~E.~Shaposhnikov,
      {\sl ``A lattice Monte-Carlo study of the hot
      electroweak phase transition"},
      CERN-TH.6901/93.}
\refitem{23}
      {W. Buchm\"uller, Z. Fodor, T. Helbig and D. Walliser,
      {\sl ``The weak electroweak phase transition"},
      DESY preprint 93-021.}
\refitem{24}
      {A.~I.~Bochkarev, S.~V.~Kuzmin and M.~E.~Shaposhnikov,
      {\sl ``Model dependence of the cosmological upper
      upper bound of the Higgs-boson mass"},
      \pr{D43}{1991}{369}.}
\refitem{25}
      {K.~Enqvist,  J.~Ignatius, K.~Kajantie and K.~Rummukainen,
      ``{\sl Nucleation and bubble growth in a first-order
      cosmological electroweak phase transition }'',
      \pr{D45}{1992}{3415}.}
\refitem{26}
      {P.~Elmfors,
        {\sl ``Finite temperature renormalization for the
       $(\phi^3)_6$-
        and $(\phi^4)_4$-models at zero momentum"},
        \zp{C56}{1992}{601};
      N.~Tetradis and C.~Wetterich,
      {\sl ``The high temperature phase
      transition for $\phi^4$-theories"},
      \np{B398}{1993}{659}.}

\printreference

\vfill\eject
\def\jumpp{\jump\noindent}
\centerline{\subsection Figure Captions}\jumpp
{\bf Fig.\ 1.} Schematic finite temperature dispersion relations for fermions:
(a) left handed particle and
right handed hole; $k_1$ marks the maximum difference between the particle and
hole energies; (b)
left handed hole and right handed particle; $k_2$ denotes the crossing point
where the energies are equal.
\jumpp
{\bf Fig.\ 2.} The self--energy diagram for the fermionic thermalization rate.
The arrows indicate the flow of four-momenta.
\jumpp
{\bf Fig.\ 3.} The self--energy diagram for the gauge bosonic thermalization
rate. The arrows indicate the flow of four-momenta.
\jumpp
{\bf Fig.\ 4.} Two two--loop diagrams contributing to the thermalization rate:
(a) pure scalar "rising sun"; (b) "rising sun" with gauge bosons.
\jumpp
{\bf Fig.\ 5.} One-loop thermalization rates: SU(2) gauge boson contribution
(dotted line),
U(1) gauge boson contribution (dashed line) and top quark contribution for the
absorption channel (a) and the decay channel (b) (solid lines).

\vfill\eject

\null
\nopagenumbers
\vskip -20pt
\line {\hfill NORDITA--93/48  P}
\vskip 1.2truecm
\centerline {\section Thermalization of the Higgs field }
\vskip 0.5truecm
\centerline{\section at the electroweak phase transition}
\vskip 1.5truecm
\centerline {Per Elmfors{\footnote{\bx $^\dagger$}{\baselineskip=9pt\bx
internet: elmfors@nordita.dk}}, Kari Enqvist{\footnote{\bx
$^\ddagger$}{\baselineskip=9pt\bx
internet: enqvist@nbivax.nbi.dk}}
and Iiro Vilja{\footnote{\bx $^\star$}{\baselineskip=9pt\bx
internet: iiro@nordita.dk}}}
\vskip 0.4truecm
\centerline {Nordita, Blegdamsvej 17, DK-2100 Copenhagen \O , Denmark}
\vskip 0.4truecm
\vskip 0.4truecm
\centerline {\bf Abstract}
\vskip 1truecm
\noindent
The thermalization rate for long wavelength fluctuations in the Higgs
field is calculated from the imaginary part of the finite temperature
effective action in the unbroken phase of the Standard Model. We use improved
propagators including a resummation of hard thermal loops.
The thermalization rate is computed at one loop level, but
an estimate of the two--loop contribution appears to give an indication that
they
are comparable to the one--loop result for small
thermal higgs mass. We show also that
the Higgs field fluctuations are likely to thermalize very fast compared with
the electroweak phase transition time.
\vfill\eject

\end